\documentclass[fleqn,usenatbib]{mnras}

\usepackage{fix-cm}
\usepackage{newtxtext,newtxmath}

\usepackage[T1]{fontenc}
\usepackage{caption}
\DeclareRobustCommand{\VAN}[3]{#2}
\let\VANthebibliography\thebibliography
\def\thebibliography{\DeclareRobustCommand{\VAN}[3]{##3}\VANthebibliography}
\usepackage{float}

\usepackage{graphicx}	
\usepackage{amsmath}	
\usepackage{float}
\usepackage{placeins}
\usepackage{orcidlink}

\title[Low Luminosity AGN 4DE1]{Do Low-Mass, Low-Luminosity AGNs Deviate from the Quasar Main Sequence?
}

\author[Sharma et al. ]{
Himanshu Sharma\orcidlink{0009-0006-8185-7322}$^{1}$\thanks{E-mail: himanshu4gya@gmail.com },
Vivek Kumar Jha\orcidlink{0000-0002-3277-6335}$^{2}$\thanks{E-mail: vivekjha.aries@gmail.com},
Hum Chand\orcidlink{0000-0002-3163-4941}$^{1}$,
Swayamtrupta Panda\orcidlink{0000-0002-5854-7426}$^{3}$\thanks{Gemini Science Fellow}
\\
$^{1}$Department of Physics and Astronomical Science, Central University of Himachal Pradesh, Dharamshala, 176215, India\\
$^{2}$National Centre for Radio Astrophysics, Tata Institute of Fundamental Research, Post Bag 3, Ganeshkhind, Pune, 411007; India\\
$^{3}$ International Gemini Observatory/NSF NOIRLab, Casilla 603, La Serena, Chile\\
}

\date{Accepted XXX. Received YYY; in original form ZZZ}

\pubyear{\the\year{}}

\begin{document}
\label{firstpage}
\pagerange{\pageref{firstpage}--\pageref{lastpage}}
\maketitle


\begin{abstract}
We present a comprehensive spectroscopic and variability-based characterisation of a sample of low-luminosity active galactic nuclei (AGNs) hosting low mass black holes, identified by $H\beta$ full width at half maximum (FWHM) $< 2200$\,km\,s$^{-1}$. While the narrow line widths are consistent with the formal definition of narrow-line Seyfert 1 (NLSy1) galaxies, the broader accretion and emission properties reveal key distinctions. The sample exhibits sub-Eddington accretion rates (median $\log R_{\mathrm{Edd}} \approx -0.68$) and comparatively weak Fe~{\sc ii} emission (median $R_{\rm Fe~{\textsc {ii}}} \approx 0.61$), in contrast to the strong Fe~{\sc ii} strengths and high Eddington ratios characteristic of classical NLSy1s. Optical variability amplitudes, derived from Zwicky Transient Facility (ZTF) light curves, are similar to those typically seen in Seyfert 1 galaxies, with a median $\log(\sigma) \approx -0.68$, suggesting the AGN component's significant contribution to variability. In the  optical plane of the 4D Eigenvector 1 (4DE1) parameter space, these sources occupy a distinct locus in the low-$R_{\rm Fe~{\textsc {ii}}}$, low-$R_{\mathrm{Edd}}$ regime, suggesting a physically distinct accretion state. Our findings indicate that this population may represent a low-accretion analogue within the broader narrow-line AGN family, offering new insights into black hole growth at low mass scales.
\end{abstract}




\section{Introduction}


Active galactic nuclei (AGNs) are extremely luminous and most variable objects in the Universe powered by gas and dust accreting onto a supermassive black hole (SMBH) at the galaxy’s centre. These systems radiate across the entire electromagnetic spectrum and are classified based on the orientation with respect to the observer, presence or absence of radio emission and a jet \citep{antonucci_1993, urry_padovni_1995, heckman_best_2014, padovani_2016}. Nevertheless, much of this diversity can be understood as the consequence of a small number of governing parameters: black hole mass, accretion rate, orientation relative to the observer, and the presence or absence of collimated outflows \citep{borson_greene_1992, urry_padovni_1995, sulentic_2000, shen_ho_2014,netzer_2015}.

Efforts to find unity in this diversity have long pursued a unified model of AGNs. A seminal study by \citet{borson_greene_1992} applied Principal Component Analysis (PCA) to 87 quasars from the Palomar-Green Bright Quasar Survey \citep{schmidt_greene_1983}, revealing that the dominant axis of variance—Eigenvector 1 (EV1)—is marked by an anti-correlation between the relative strength of optical Fe~{\sc ii} emission ($R_{\rm Fe~{\textsc {ii}}}$) and the width of the broad H$\beta$ emission line. This correlation—subsequently known as the quasar main sequence—has since become a foundational organising principle in type 1 AGN \citep{sulentic_2000, zamfir_2008, marziani_sulentic_2014, shen_ho_2014, sun_shen_2015,panda_2024}.

Expanding upon this framework, the 4D Eigenvector 1 (4DE1) formalism incorporates additional diagnostics such as the soft X-ray photon index and C~{\sc iv} emission line blueshift, thereby offering a more nuanced and multidimensional portrait of AGN phenomenology \citep{sulentic_2007, zamfir_2010, Richards_2011, Kruczek_2011, brotherton_2015}. This schema stratifies AGNs into Population A (FWHM(H$\beta$) < 4000 $km \; s^{-1}$) and Population B (FWHM(H$\beta$) > 4000 $km \; s^{-1}$, associated respectively with high and moderate accretion rates. Population A sources—often including Narrow-Line Seyfert 1 galaxies (NLSy1s)—are not only high‑Eddington systems observed at relatively low inclination,  but also preferentially host lower‑mass black holes.

Despite its successes, the 4DE1 formalism has been primarily applied to luminous, unobscured quasars. Its extrapolation to low-luminosity AGNs (LLAGNs) remains an open frontier. LLAGNs—characterised by weak emission lines, low ionisation parameters, and inefficient accretion modes—are ubiquitous in the local Universe and often inhabit late-type or elliptical galaxies with relatively low stellar mass \citep{greene_ho_2007, ho2008nuclear}. In many cases, their central black holes lie in the low-mass regime ($10^4$–$10^6,M_\odot$) \citep{dong_2012, reines_2013, baldassare_2015, koliopanos_2017, liu_2018}, occupying the largely observationally unconstrained space between stellar-mass black holes and classical SMBHs \citep{greene_ho_2007, mezcua_2017, greene2020intermediate}. The study of such systems offers a rare glimpse into the mechanisms of SMBH seed formation, the long tail of AGN duty cycles, and the low-luminosity limit of AGN feedback on host galaxies \citep{reines_2013, graham_2016, reines_2020, 2025arXiv250314745D}. Recent studies such as \citet{panda_2024} have revisited the quasar main sequence framework, reinforcing the significance of EV1 as a diagnostic axis potentially rooted in accretion physics and orientation \citep{highz_2023_qms, marziani2024metallicity} . However, the study of low-Eddington and low-luminosity sources remains minimal, leaving the applicability of this framework across the full AGN luminosity function unresolved.

    Given the mass-luminosity scaling relation, black holes in the $10^4 - 10^6 M_\odot$  range will naturally produce lower bolometric outputs than their supermassive counterparts, making LLAGNs ideal candidates for harbouring such systems \citep{volonteri_2010, greene2020intermediate}. Furthermore, LLAGNs are often found in late-type or dwarf galaxies with relatively low stellar masses \citep{reines_2013, baldassare_2015, mezcua_2017}. These galaxies tend to have experienced a quieter evolutionary history with fewer major mergers, which curtails the rapid growth typically required to form supermassive black holes \citep{volonteri_2009, koliopanos_2017, greene2020intermediate}. Consequently, their central black holes remain in the low-mass regime.

From a theoretical standpoint, it remains to be clarified whether LLAGNs represent a fundamentally distinct class of accretors or whether they occupy a continuous position along the 4DE1 axis, albeit shifted due to reduced radiative efficiency and lower Eddington ratios \citep{marziani_2001, ho2008nuclear}. Crucially, if the quasar main sequence and 4DE1 are not merely taxonomical tools but encode evolutionary pathways—driven by SMBH growth, host galaxy evolution, and accretion history—then LLAGNs may either represent early seed phases or the faded remnants of formerly active quasars \citep{hickox_2009, shankar_2013}. 



Using optical spectra from the Sloan Digital Sky Survey (SDSS) \citep{york_2000}, we measure Fe~{\sc ii} strength, H$\beta$ FWHM, and continuum luminosity to place LLAGNs within the 4DE1 framework, probing whether their spectral properties resemble those of luminous quasars or suggest unique accretion and host galaxy interactions \citep{ho2008nuclear}. In addition to spectral diagnostics, variability provides a complementary probe of accretion physics in LLAGNs. Optical variability on timescales of weeks to months has been successfully modeled using a Damped Random Walk (DRW) process, with the variability amplitude and damping timescale serving as empirical tracers of accretion disk fluctuations \citep{Kelly2009, MacLeod2010, Zu2013}. For low-luminosity sources, however, host galaxy contamination can significantly dilute the intrinsic AGN signal, making it essential to account for the AGN fraction when interpreting variability properties. Including both spectral and variability diagnostics therefore allows a more complete assessment of whether LLAGNs conform to the 4DE1 framework or reveal distinct behavior shaped by their low Eddington ratios.

Our objectives are twofold: (1) to investigate whether the 4DE1 formalism, established for luminous AGNs, extends to LLAGNs or requires rethinking due to their low-luminosity regime, and (2) to examine whether LLAGNs trace an evolutionary path linked to the quasar main sequence, potentially as faded remnants of luminous quasars or early seeds of supermassive black hole growth \citep{reines_2020}. These results will be helpful to understand the physics of low-accretion systems and their role in black hole and galaxy co-evolution \citep{2013ARA&A..51..511K}.

This paper is structured as follows. In Section \ref{section2}, we describe the sample selection and the criteria used for source inclusion. Section \ref{section3} outlines the spectral fitting procedure and the methods employed for estimating emission line parameters. The results, including the derived spectral properties, parameter distributions, and their mutual correlations, are presented in Section \ref{section4}. Section \ref{section5} provides a discussion of these findings. Finally, the conclusions are summarized in Section \ref{section6}.

\section{sample} \label{section2}
\begin{figure}
    \centering
        \includegraphics[width=.52\textwidth]{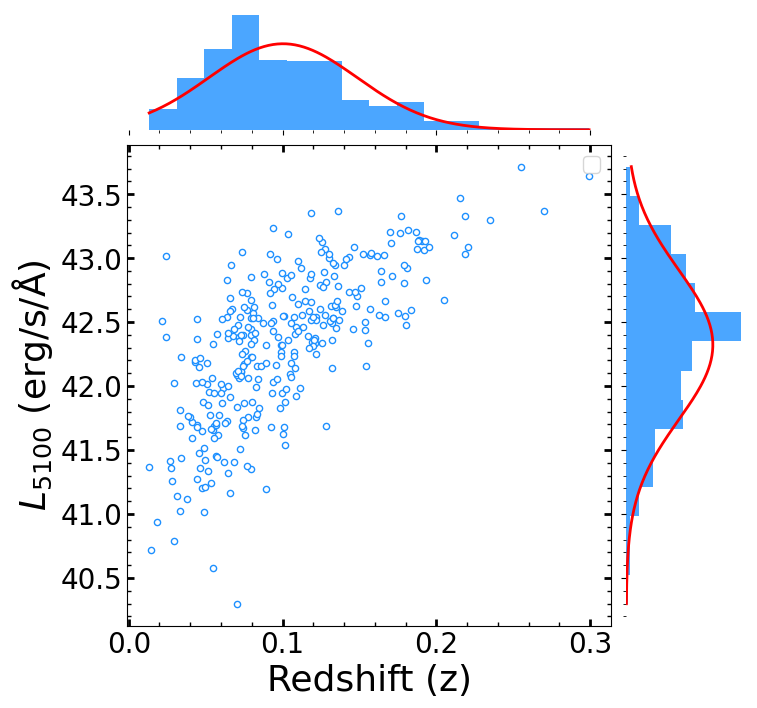} 
        \caption{The distribution of continuum luminosity at 5100 Å ($L_{5100}$) as a function of redshift ($z$) for our final sample. The marginal histograms along the axes show the one–dimensional distributions of $L_{5100}$ and $z$.}
       
        \label{fig: l_z_plot}
\end{figure}

We constructed a sample of 513 unique low-luminosity active galactic nuclei (LLAGNs) hosting  low mass black holes by combining catalogues from \citet{dong_2012} and \citet{liu_2018}, based on the Sloan Digital Sky Survey (SDSS) Data Releases 4 and 7, respectively. All sources have redshifts $z < 0.35$, with a median redshift of 0.09 and their redshift and luminosity distributions at 5100~\AA\ are shown in Figure~\ref{fig: l_z_plot}.

\begin{figure*}
    \centering
        \includegraphics[width=\textwidth]{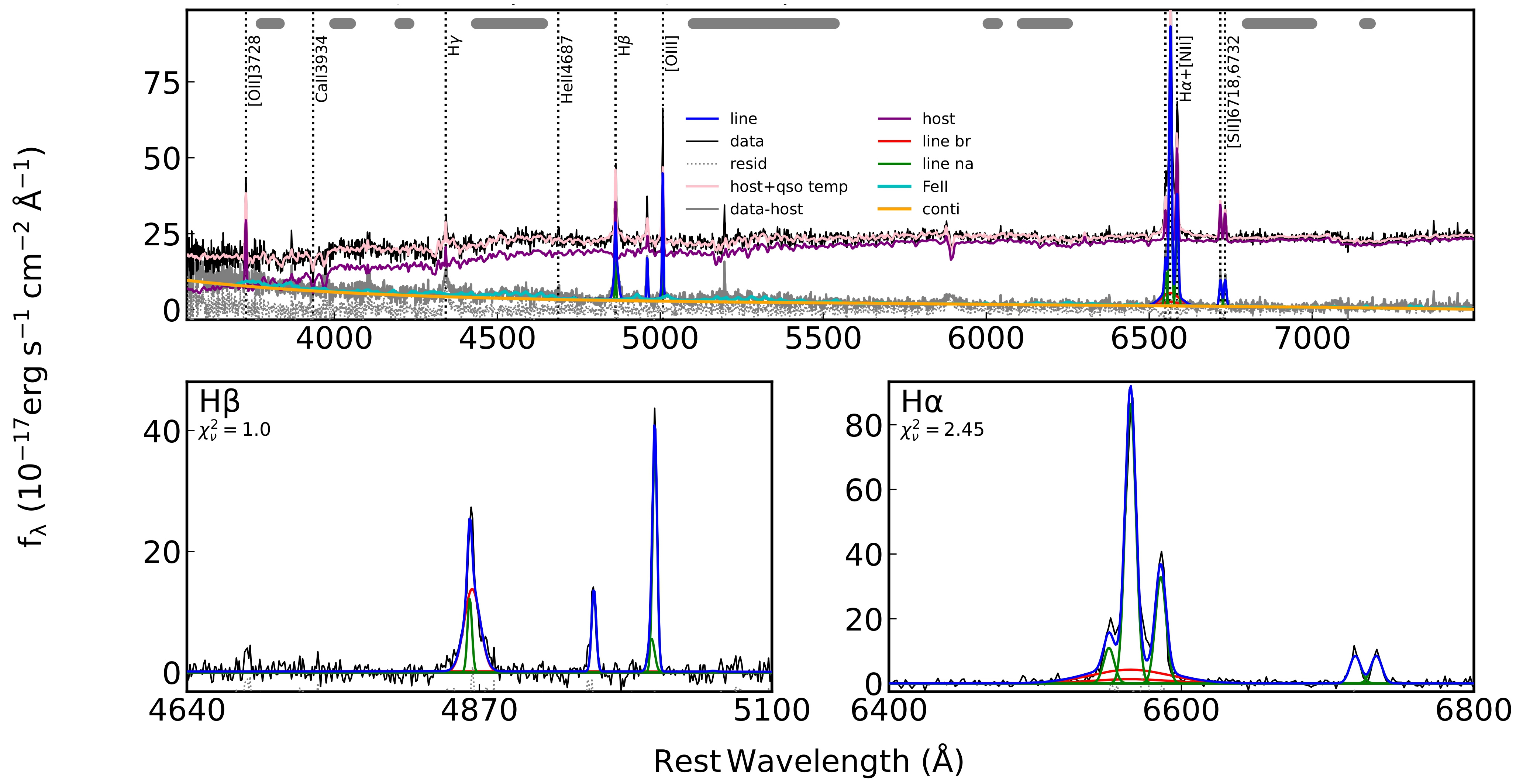} 
        \caption{An illustration of a good fit from \texttt{PyQsoFit} for the source J135159.16+025034.0, {Top panel} represents the host galaxy and iron template plus continuum subtracted fitted spectrum along with continuum and individual templates of iron and host galaxy represented by different colours. \emph{Bottom left} shows the $H\beta$ fitted region and \emph{bottom right} shows the $H\alpha$ fitted region. The blue line represents the emission line model, and the red lines represent the multi-Gaussians used to build the final emission line model. For fitting broad emission lines, we have used single Gaussian with the constraint of FWHM > 700 $km \; s^{-1}$, while for narrow line regions, we have used a single Gaussian with FWHM < 700 $km \; s^{-1}$. For fitting the tails of broad $H\beta$ line, we have also used a very broad single Gaussian with FWHM > 10,000 $km \; s^{-1}$ without assigning any physical meaning to it.}
       
        \label{g_spec}
\end{figure*}

We retrieved optical spectra for all 513 sources from SDSS Data Release 16 \citep{sdss_dr14q_cat}, which provides updated data for sources identified initially in earlier releases. For sources with multiple spectra observed across different epochs, we selected the spectrum with the highest signal-to-noise ratio (S/N) to ensure optimal data quality. We then applied an initial quality cut, retaining only spectra with an S/N $\geq 10$ in the continuum near the H$\beta$ emission line, excluding 61 sources and a refined sample of 452 LLAGNs.

Spectral analysis was performed using the open-source Python-based software \texttt{PyQSOFit  v2.1.6} \citep{guo_2018}\footnote{\href{https://github.com/legolason/PyQSOFit}{PyQSOFit GitHub Repository}}, which models the continuum, Fe~{\sc ii} emission, and emission lines to derive key parameters for the 4D Eigenvector 1 (4DE1) framework  \citep{sulentic_2000,richards_2006}. 
 

Out of the 513 sources in our parent sample, acceptable spectral fits could be achieved for 449 objects, with an illustrative example shown in Figure \ref{g_spec}.
Additionally, we show examples of sources where our fitting process fails, either due to low S/N or because the H$\beta$ emission line is not clearly detected (see Figures  \ref{fig:a1} and \ref{fig:a2} in the Appendix).
Given that low mass AGNs typically have black hole masses one order of magnitude lower than those in Narrow-Line Seyfert 1 (NLSy1) galaxies \citep{greene_ho_2007}, we expect their broad-line regions (BLRs) to be compact, resulting in H$\beta$ FWHM values of a few thousand km s$^{-1}$. To ensure consistency with the low luminosty AGN mass range, we applied an additional constraint of H$\beta$ FWHM $< 2200$ km s$^{-1}$, which excluded 134 sources with unusually broad lines, likely indicative of more massive black holes or fitting errors. As a result, the final sample comprises 315 LLAGNs with reliable spectral fits, suitable for analysis within the 4DE1 framework. 

\section{Analysis} \label{section3}


As mentioned earlier, we employed the Python-based spectral fitting code \texttt{PyQSOFit} to extract the properties of the emission lines. The fitting procedure involves several steps, which we briefly outline here. First, each spectrum is shifted to the rest frame of the source. The continuum is then modeled using a combination of a power-law, a low-order polynomial, an Fe~{\sc ii} emission template, and a host-galaxy template. The Fe~{\sc ii} templates are taken from \citet{borson_greene_1992}, while the host-galaxy component is derived from \citet{yip_2004_host_gal_temp} using principal component analysis (PCA). After subtracting the best-fit continuum (power law + polynomial + Fe~{\sc ii} + host galaxy), the residual spectrum primarily contains the emission-line features. These lines are subsequently fitted with multiple Gaussian components, with narrow Gaussians representing emission from the narrow-line region and broad Gaussians accounting for the broad-line region.

We primarily focused on fitting $H\beta$ line in the wavelength range of 4640.0 \AA \; to 5100 \AA.  To model the $H\beta$ region, we used a single Gaussian for the broad component ($FWHM > 700$ $km \; s^{-1}$) and a single Gaussian for the narrow component ($FWHM \leq 700$ $km \; s^{-1}$). When extended wings were present, we added an additional very-broad Gaussian ($FWHM > 10000$ $km \; s^{-1}$) to reproduce the far-line tails without assigning any physical meaning to it. Crucially, we constrained the broad $H\beta$ width by tying it to the optical Fe~{\sc ii} template broadening obtained from the continuum fit, implemented as a tight $(\pm 10\%)$ window around the Fe~{\sc ii} derived value to allow small deviations (e.g., see \citet{tie_hbeta_fe2}).


The mass of SMBH was estimated using the FWHM of the broad component of the $H\beta$ emission line from a single epoch spectra using the virial relation \citep[e.g., see][]{Kaspi_2000, greene_ho}. Example of our best fit $H\beta$ and $H\alpha$ is illustrated in bottom left and right panel of Figure \ref{g_spec} respectively.

We also estimated the Eddington ratio from the fitted parameters. It is defined as the ratio between bolometeric luminosity to Eddington luminosity $L_{bol}/L_{\text{Edd}}$ where $L_{bol}$ is calculated from the estimated continuum luminosity at 5100 \AA~ as $L_{bol} = 5 \times L_{5100}$ and $L_{\text{Edd}}$ is calculated from the estimated BH mass as $1.3 \times 10^{38} \, (M_{BH}/M_\odot) \, erg \; s^{-1}$. We note that the canonical quasar bolometric correction of $L_{\rm bol}=9.1\,L_{5100}$, derived from the SDSS composite SED of luminous quasars, is not strictly applicable to the low-luminosity regime considered here. High–spatial resolution studies demonstrate that LLAGN SEDs differ markedly from those of classical Seyferts and quasars, exhibiting a suppressed Big Blue Bump and enhanced IR and X-ray contributions \citep{Fernandez-Ontiveros2012_SEDLLAGN}. As a result, the optical continuum constitutes a larger fraction of the total radiative output, implying a smaller optical bolometric correction. Motivated by the broadband AGN SED analysis of \citet{2020A&A...636A..73D}, we therefore adopt a conservative correction of $L_{\rm bol}=5\,L_{5100}$, which is more appropriate for LLAGNs than the canonical quasar value.

Another key parameter in the context of the Quasar Main Sequence that we have estimated is the Fe~{\sc ii} strength, denoted as $R_{\rm Fe~{\textsc {ii}}}$. The parameter $R_{\rm Fe~{\textsc {ii}}}$ is defined as the ratio of the equivalent widths of Fe~{\sc ii} and that of H$\beta$ broad component: $R_{\rm Fe~{\textsc {ii}}}$ = EW(Fe~{\sc ii})/EW(H$\beta$). The equivalent width \textit{W} of an emission line is given by
\begin{equation}
    W = \int \frac{(F_c - F_\lambda)}{F_c} d\lambda
\end{equation}
where $F_c$ is the continuum flux and $F_\lambda$ is the observed flux at wavelength $\lambda$ within the integration interval. For Fe~{\sc ii}, the equivalent width was measured over the wavelength range $4435 - 4685$ \AA , while for $H\beta$, the integration range was $4640 - 5100$ \AA $\,$  \citep{borson_greene_1992, Marziani_2003, yu_shen_2019}. Since the spectra were shifted to the rest frame prior to fitting in PyQSOFit, all equivalent widths reported in this work correspond to rest-frame values.

For estimation of the black hole mass of each of the sources, we used virial estimators for which we need FWHM of the emission lines; the most commonly used line is $H\beta$ as this line has been calibrated as the virial estimator of black hole mass \citep{Vestergaard_2002, VP_2006, Wang_2009, netzer_2012}. Furthermore, the mass-scaling relation based on the $H\beta$ line luminosity provides a robust alternative for estimating black hole masses in low-redshift AGNs, where the optical continuum is often significantly contaminated by host galaxy starlight or nonthermal emission \citep{VP_2006}. It is important to highlight that the SMBH masses in the samples of \citet{greene_ho, greene_ho_2007, dong_2012, liu_2018} were estimated using a virial method based on the broad $H\alpha$ line width. However, fitting the $H\alpha$ line is challenging due to persistent blending with the N~{\sc ii} lines. Therefore, in our analysis, we instead estimate the black hole mass using the FWHM of the broad component $H\beta$ line as done for other type -1 AGN in the literature.

\begin{figure*} 
    \centering
        \includegraphics[width=0.33\textwidth]{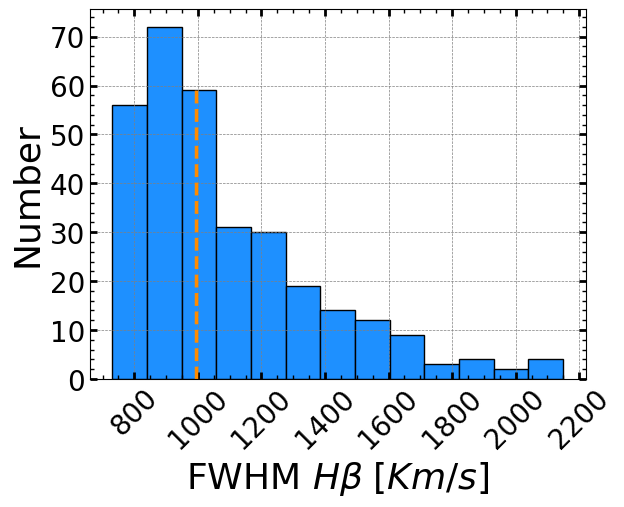} 
        \includegraphics[width=0.33\textwidth]{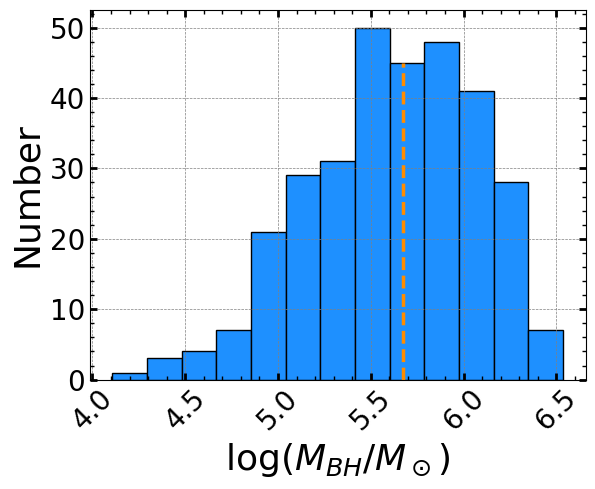}
        \includegraphics[width=0.33\textwidth]{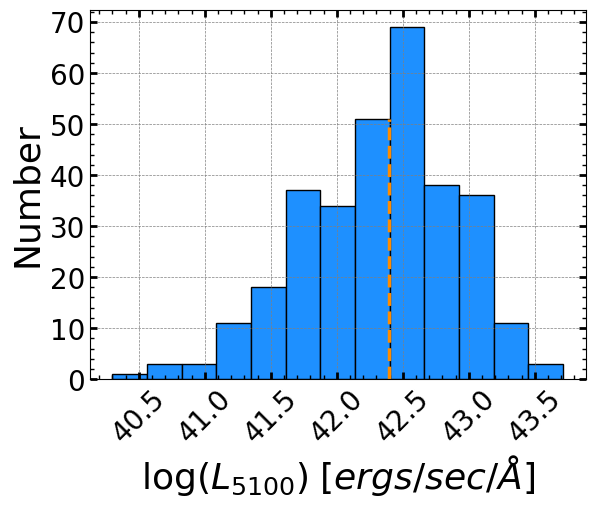}

        \caption{Distributions of FWHM $H\beta$, log($M_{BH}/M_\odot$) and log($L_{5100}$) for the final sample of 315 LLAGNs with red dashed lines mark the median values: 992.42 $km \; s^{-1}$, 5.67 and 42.40 respectively.}

        \label{fig: histograms_fig3}
\end{figure*}

\begin{figure}
        \includegraphics[width=0.44\textwidth]{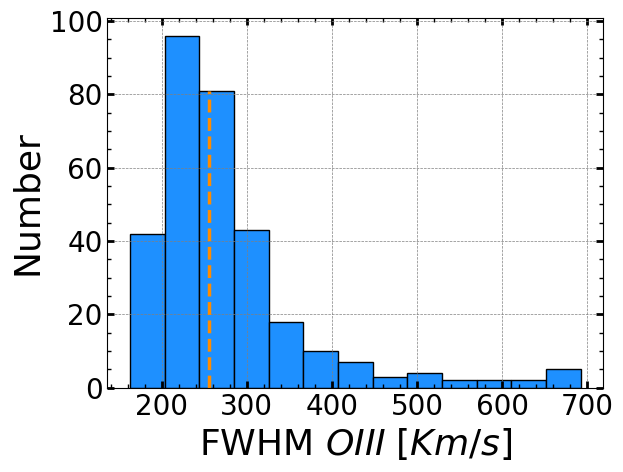}
        \includegraphics[width=0.44\textwidth]{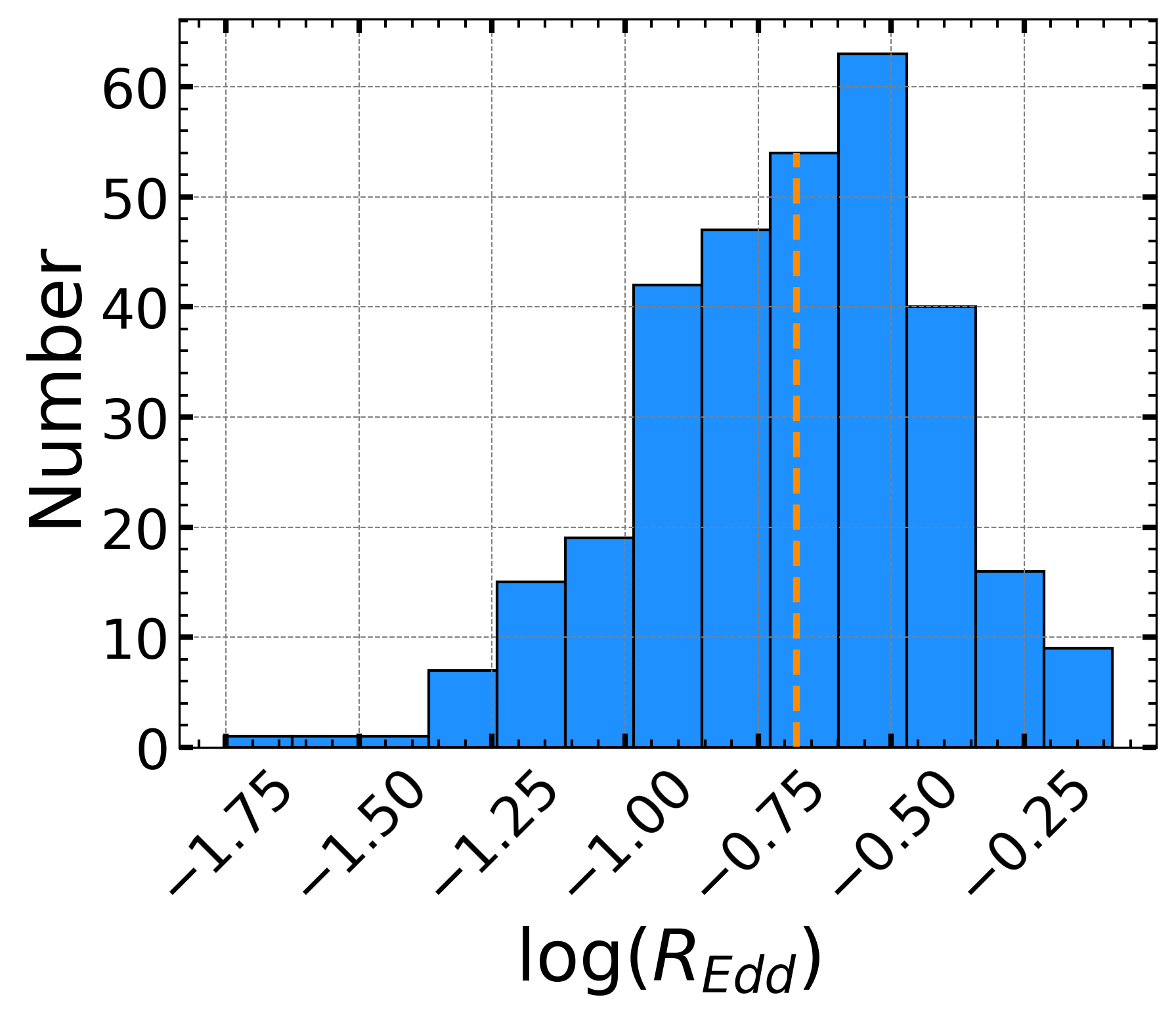}
        \includegraphics[width=0.44\textwidth]{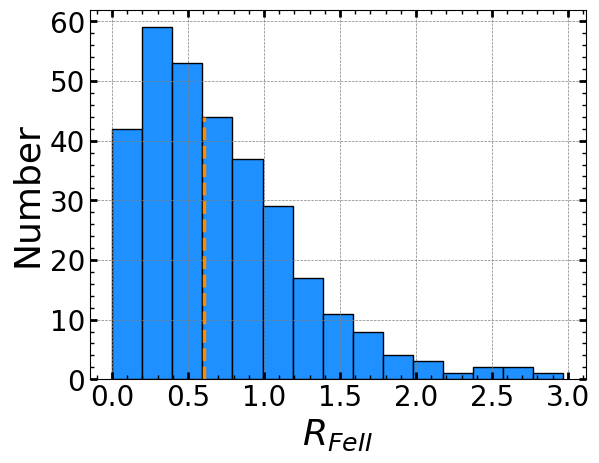}
        
        \caption{Distributions of parameters derived from spectral emission-line fitting, with median values of 255.54 $km \; s^{-1}$, –0.68, and 0.61 respectively, marked by red dashed lines. For visualisation, the $R_{{\mathrm{Fe~\textsc{ii}}}}$ histogram is shown with a cut at $R_{{\mathrm{Fe~\textsc{ii}}}} < 3$, excluding two sources, while the other parameters are plotted for the full sample.}
       
        \label{fig: histograms_fig4}
\end{figure}

\section{Results} \label{section4}

\begin{figure} 
    \centering
        \includegraphics[width=0.5\textwidth]{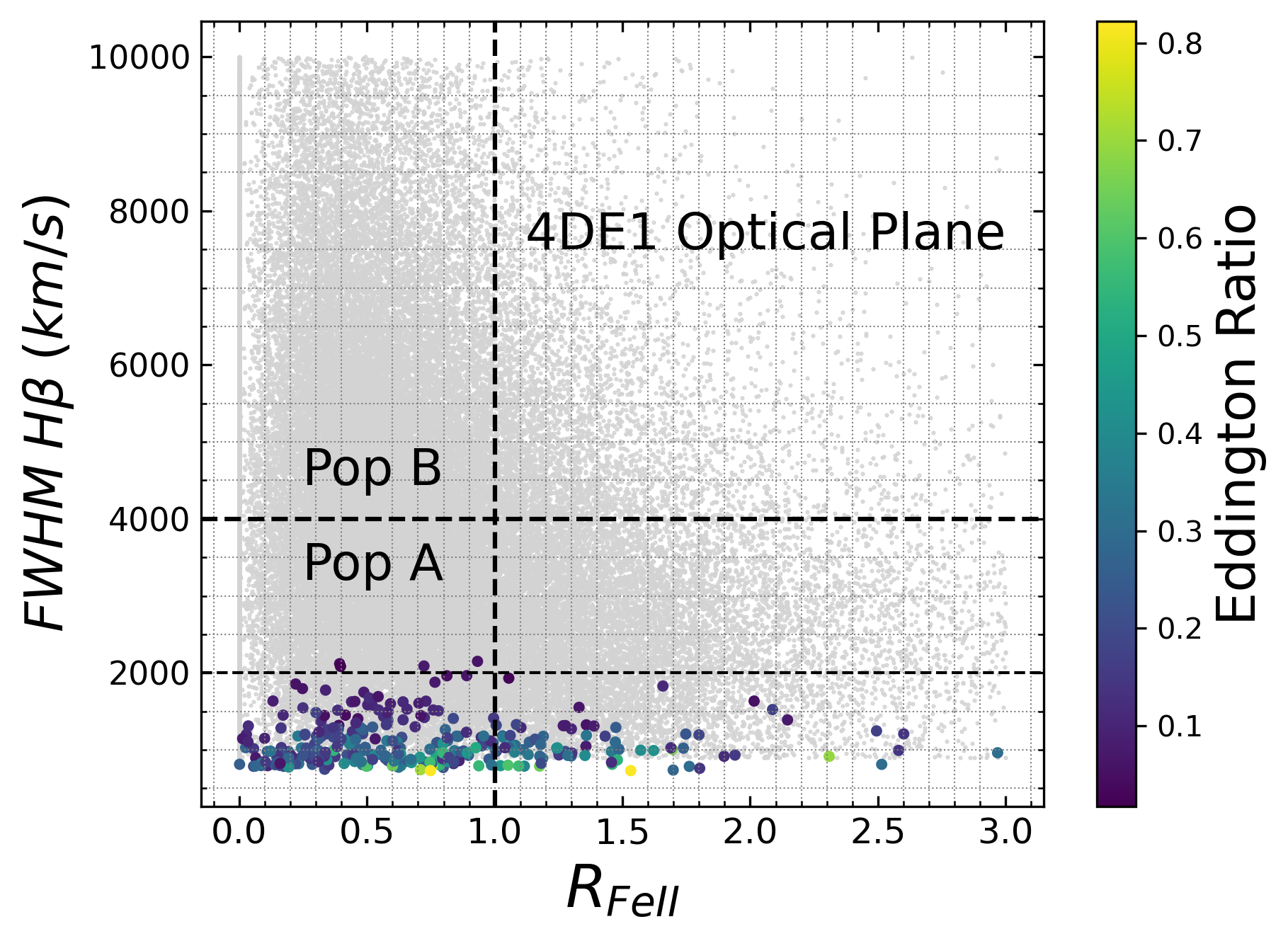}
        \caption{Correlation plot between FWHM of $\;H\beta$ and   $R_{\rm Fe~{\textsc {ii}}}$ constituting the quasar main sequence(QMS). 
        QMS plot from our sample of IMBH. The coloured dots represent the IMBH sources while the semi-transparent grey points indicate the general main sequence of quasars taken from \citet{sdss_dr14q_cat}. The thick horizontal Black dashed line at $4000$ $km \; s^{-1}$ separates the two different populations named Pop A and Pop B, while the thin horizontal black dashed at $2000$ $km \; s^{-1}$ indicates the limit of NLSy1s. Thick vertical Black dashed line at $R_{\rm Fe~{\textsc {ii}}} = 1$  separates the extreme Population A to the right. $R_{\rm Fe~{\textsc {ii}}}$ is truncated to < 3, with the exclusion of 2 sources, for better visualisation.} 
        \label{fig: Main_seq.}
\end{figure}

\begin{table*}
\centering
\resizebox{\textwidth}{!}{%
\begin{tabular}{ccccccccccc}
\hline
SDSS Name & $z$ & FWHM$_{H\beta}$ &  EW$_{\rm Fe~{\textsc {ii}}}$ & $\log L_{5100}$ & Host Fraction & $\log M_{\rm BH}$ & $R_{\rm Fe~{\textsc {ii}}}$ & $\log R_{\rm Edd}$ & $R_{\rm BLR}$ & $\log L_{\rm bol}$ \\
 & & [$km \; s^{-1}$] & [\AA] & [erg/s] & [\%] & [$M_\odot$] &  &  &[lt-days] & [erg/s] \\
\hline
J005633.36+000510.2 & 0.08 & 1403.71 ± 22.26 & 140.58 & 41.35 ± 0.05 & 95 & 5.24 ± 0.01 & 0.46 & -1.30 ± 0.01 & 0.04 ± 0.00 & 42.30 ± 0.05 \\
J010712.04+140845.0 & 0.08 & 906.31 ± 6.24 & 22.97 & 42.59 ± 0.01 &  68 & 5.66 ± 0.01 & 0.15 & -0.47 ± 0.01 & 0.76 ± 0.01 & 43.55 ± 0.01 \\
J012150.64-100510.0 & 0.10 & 808.81 ± 17.86 & 41.23 & 42.36 ± 0.01 & 90 & 5.41 ± 0.02 & 0.48 & -0.44 ± 0.02 & 0.45 ± 0.01 & 43.32 ± 0.01 \\
...        & ... & ... & ... & ... & ... & ... & ... &... & ... & ... \\
\hline
\end{tabular}%
}
\caption{The full catalogue is available in the online version of this paper. This portion of the table is shown for guidance regarding its format and content.
}
\label{tab:sample}
\end{table*}

\subsection{Spectral Properties and Derived Quantities}

The key physical properties of our AGN sample have been estimated using optical spectra. Directly measurable quantities include the full width at half maximum (FWHM) of the broad H$\beta$ emission line and the continuum luminosity at 5100\,\AA\ ($L_{5100}$). These are used to compute the black hole mass ($M_{\mathrm{BH}}$) via the standard virial method.
The bolometric luminosity ($L_{\mathrm{bol}}$) is estimated by applying a bolometric correction factor of 5 to $L_{5100}$. 

The Eddington ratio ($R_{\mathrm{Edd}}$), serves as an indicator of accretion efficiency and evolutionary state. The strength of optical Fe~{\sc ii} emission, quantified by $R_{\rm Fe~{\textsc {ii}}}$, provides a measure of BLR physical conditions. 

\subsection{Distributions of Parameters}

Figure \ref{fig: histograms_fig3} \& \ref{fig: histograms_fig4} shows the distributions of FWHM H$\beta$, $M_{\mathrm{BH}}$, $L_{5100}$, FWHM O~{\sc iii}, $R_{\mathrm{Edd}}$, $R_{\rm Fe~{\textsc {ii}}}$. The median values are calculated for the full sample defined by the H$\beta$ FWHM $<2200$ $km \; s^{-1}$ criterion. For visualization purposes, only the $R_{\rm Fe~{\textsc {ii}}}$ histogram is truncated with values restricted to $R_{\rm Fe~{\textsc {ii}}} < 3$ to suppress the long tail.

The median FWHM of the H$\beta$ line is  $990.75\pm20.80$ $km \; s^{-1}$, consistent with expectations for low-mass AGNs, and lower than the typical broad-line Seyfert 1 (BLSy1) AGNs. This lies within the range of narrow-line Seyfert 1 (NLSy1) galaxies, defined as systems with FWHM(H$\beta$) < 2000 $km \; s^{-1}$ \citep{goodrich_nlsy_2000km/s}. The median continuum luminosity at 5100 \AA\ is $\log L_{\mathrm{5100}} = 42.39\pm0.02$ (erg/s), roughly 1–2 orders of magnitude below that of typical NLSy1s \citep{vivek_nlsy_blsy}. 

Using the virial method, we find a median black hole mass of $\log (M_{\mathrm{BH}}/M_\odot) = 5.66 \pm 0.02$, again significantly lower than that of NLSy1s and BLSy1s, whose black hole masses typically range between $10^6$–$10^8$\,M$_\odot$. The median Eddington ratio is found to be $\log R_{\mathrm{Edd}} = -0.68$, also lower than values typically reported for NLSy1s \citep{vivek_nlsy_blsy, paliya_cat_nlsy}. Interestingly, \citet{Liu2016_lowredd_nlsy} argue that Eddington ratios may be overestimated in NLSy1s, citing uncertainties in $L_{5100}$ and line widths. The median Fe~{\sc ii} strength, $R_{\rm Fe~{\textsc {ii}}} = 0.62$, is lower than the average value of $\sim$1.0 reported for NLSy1s. 

\begin{table}
\centering
\caption{Summary of derived AGN parameters for the LLAGN sample.}
\label{tab:parameters}
\begin{tabular}{lccc}
\hline
\textbf{Parameter} & \textbf{Minimum} & \textbf{Maximum} & \textbf{Median} \\
\hline
FWHM(H$\beta$) [$km \; s^{-1}$] & 730.26 & 2149.43 & 992.42 \\
$\log L_{5100}$ [erg/s] & 40.30 & 43.71 & 42.40 \\
$\log (M_{\mathrm{BH}}/M_\odot)$ & 4.11 & 6.54 & 5.67 \\
$\log R_{\mathrm{Edd}}$ & -1.75 & -0.09 & -0.68 \\
$R_{\text{Fe\textsc{ii}}}$ & 0 & 7.22 & 0.61 \\

\hline
\end{tabular}
\end{table}

\subsection{Correlation Analysis}



We used Spearman rank tests to search for monotonic trends among the derived AGN parameters. 
A strong anti-correlation is seen between $R_{Edd}$ and the broad-line width FWHM\,$H\beta$ ($\rho=-0.57$). This is expected both empirically and by construction. Empirically, multiple SDSS-based works report an anti-correlation between $R_{Edd}$ and FWHM of the $H\beta$ emission line across NLSy1 and BLSy1 populations \citep{rakshit_cat_nlsy,vivek_nlsy_blsy}. By construction, single-epoch virial scalings give $M_{BH}\!\propto\! L^{\alpha}\,{FWHM}^2$ and hence $R_{Edd}\!\propto\!L^{1-\alpha}\,{FWHM}^{-2}$, which naturally induces an inverse dependence on FWHM . 

\begin{figure*} 
 
        \includegraphics[width=0.45\textwidth]{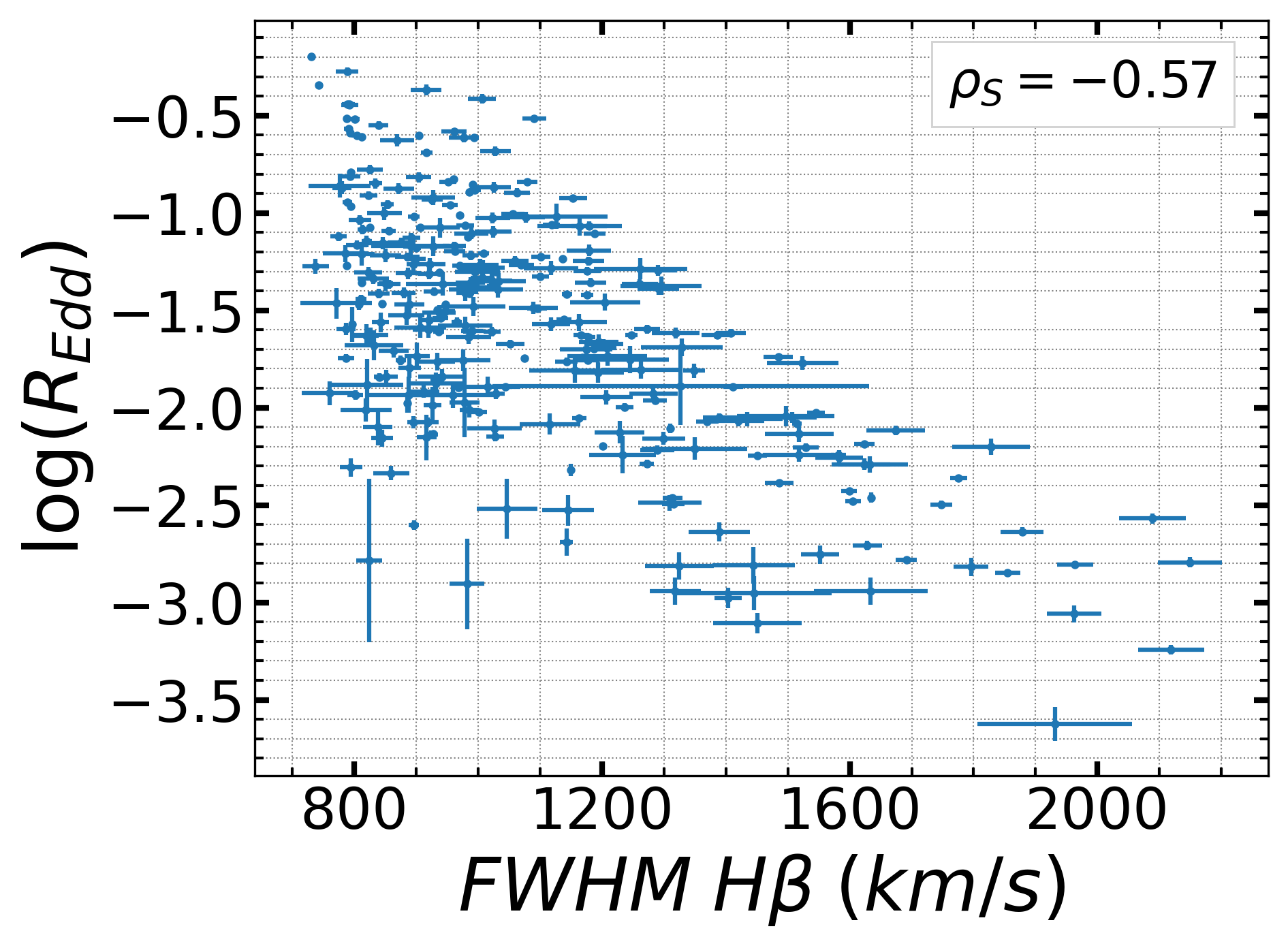}
        \includegraphics[width=0.45\textwidth]{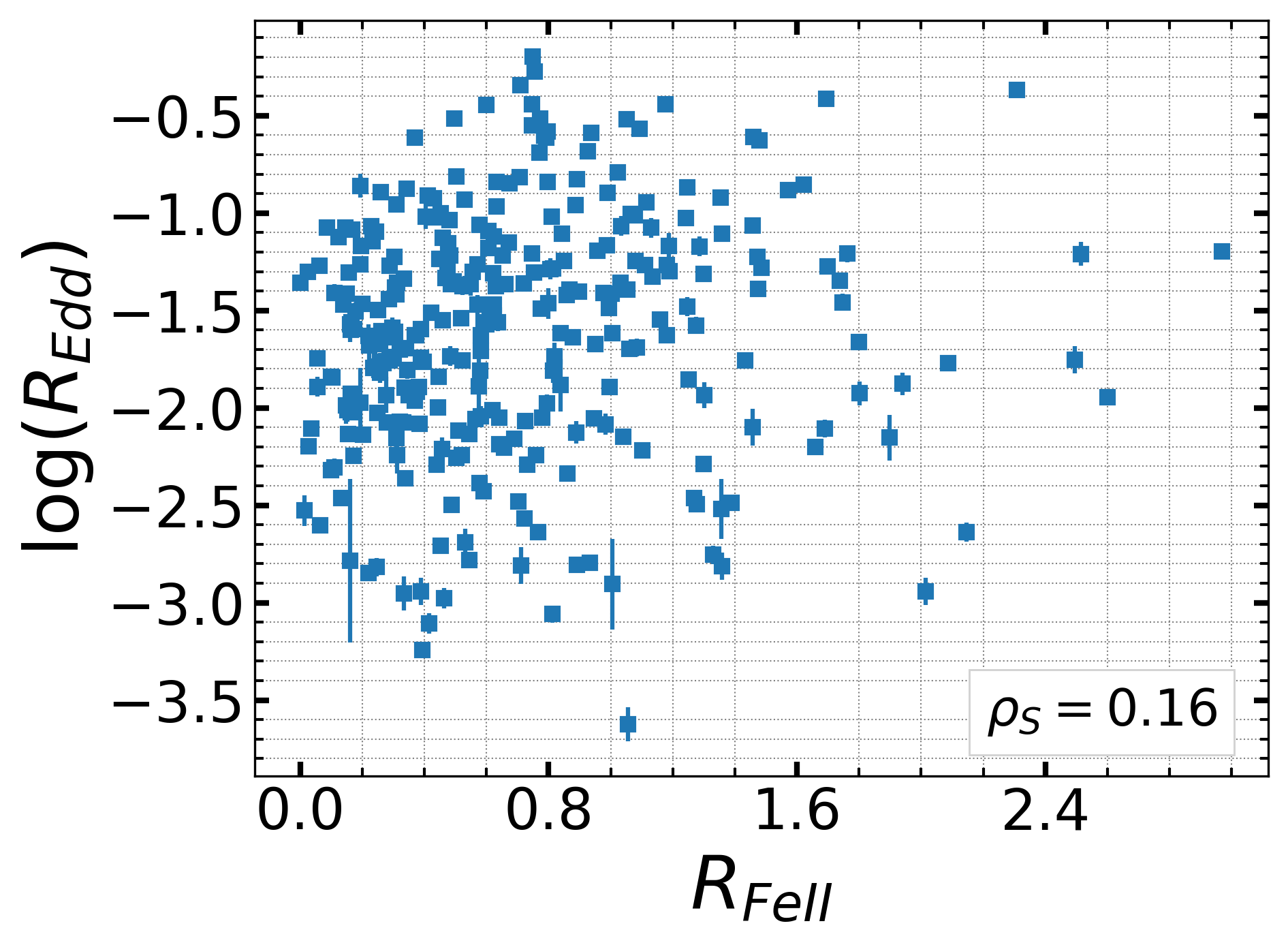}

        \caption{Correlation plots of various parameters derived from the spectral fitting procedure, along with their Spearman rank correlation coefficients. The left panel shows the correlation of the FWHM of the H$\beta$ line with $\log(R_{\mathrm{Edd}})$, while the right panel displays the correlations of $\log(R_{\mathrm{Edd}})$ with  R$_{\text{Fe~{\sc ii}}}$.} 
        \label{fig: corr_plots}
\end{figure*}

We also recover a weak positive correlation between $\log(R_{\mathrm{Edd}})$ and R$_{\text{Fe~{\sc ii}}}$ ($\rho = 0.16$), in agreement with previous studies linking Fe~{\sc ii} strength to high accretion rates \citep{rakshit_2021}. Similarly \citet{negrete_2018_corr_redd_rfe2}  reported a positive correlation between the Eddington ratio and Fe~{\sc ii} strength in a sample of 315 highly accreting sources, reinforcing the interpretation that enhanced Fe~{\sc ii} emission is a tracer of high accretion activity (see right panel of Figure~\ref{fig: corr_plots}). Finally, we find that $M_{\mathrm{BH}}$ correlates strongly with the FWHM of H$\beta$ ($\rho = 0.60$), consistent with virial expectations.

\subsection{The 4DE1 picture}


A central aspect of the quasar main sequence is the relationship between the full width at half maximum (FWHM) of the broad $H\beta$ emission line and the optical iron strength, quantified by  $R_{\rm Fe~{\textsc {ii}}}$. This “optical plane” of 4D Eigenvector 1 (4DE1) has been widely studied and organizes quasars into Population A (narrower $H\beta$, typically stronger Fe~{\sc ii}) and Population B (broader $H\beta$, weaker Fe~{\sc ii}), with Eddington ratio as the primary physical driver and orientation adding substantial dispersion in $H\beta$ width \citep{sulentic_2000, shen_ho_2014, review_marziani}. Figure~\ref{fig: Main_seq.} shows our low-mass black hole AGN sample in this plane. Interestingly, we find a very weak correlation between $R_{\rm Fe~{\textsc {ii}}}$ and $\mathrm{FWHM}(H\beta)$ ($\rho = 0.07$), suggesting that low-mass AGNs may not follow the classical 4DE1 trends seen in higher-mass, higher-luminosity quasars.

\subsection{Host Galaxy contribution and Optical variability}

\begin{figure} 
    \centering
        \includegraphics[width=0.4\textwidth]{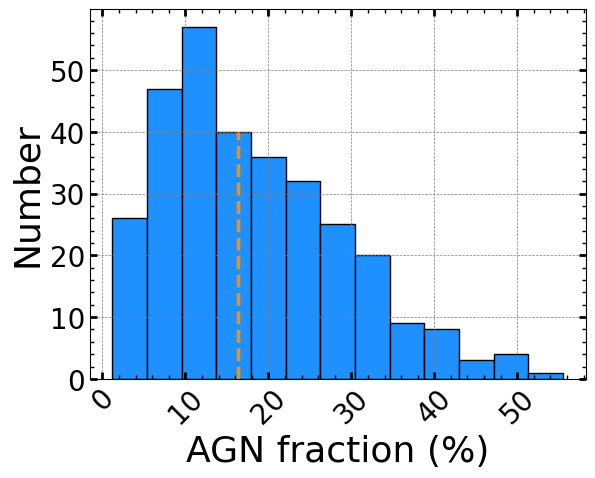}
        \includegraphics[width=0.4\textwidth]{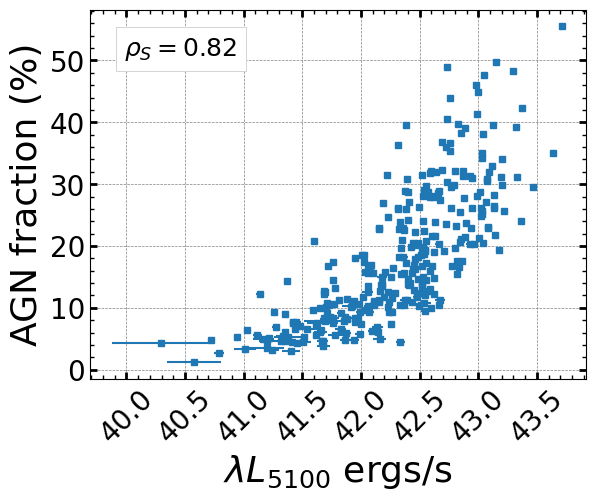}
        \includegraphics[width=0.4\textwidth]{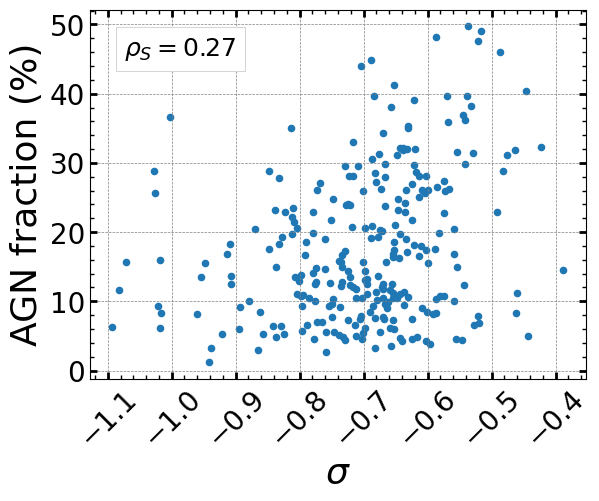}


        \caption{The top panel shows the distribution of percentge AGN fraction
        (i.e.,  100 $\times {Flux_{\textsc{agn}}}/{Flux_{\textsc{agn}+\textsc{host}}}$),
        with the median value of 16.33, marked by a vertical red dashed line. The middle panel presents AGN fraction versus $\lambda L_{5100}$, and the bottom panel shows variability amplitude ($\sigma$) versus AGN fraction. Spearman rank correlation coefficients are given for the latter two, with the positive trend in the bottom panel indicating that variability is suppressed in sources with lower AGN contribution due to host–galaxy dilution.}
        \label{fig: corr_sigma_host_frac}
\end{figure}

For low-luminosity AGNs, host galaxy contamination constitutes a significant fraction of the total optical flux. In the faintest sources, the host contribution can reach up to 90\% of the total flux at 5100\,\AA\  \citep{jalan_2023_host_frac,Ren_2024_host_frac}. As a result of low-luminosity, low-mass black hole AGNs must carefully account for host contamination, as neglecting this can lead to substantial errors in derived AGN properties such as luminosity and Eddington ratio. \texttt{PyQSOFit} provides the functionality to estimate the host galaxy fraction, and we were able to model the host component for most sources using the built-in galaxy templates (see top panel of Figure \ref{fig: corr_sigma_host_frac}).

In addition to the spectroscopic decomposition, we investigated the long-term optical variability properties of our sample using r-band Zwicky Transient Facility (ZTF) light curves. We modeled the light curves using a Damped Random Walk (DRW) process based on the studies on type 1 AGN \citep[see][for the method]{Kelly2009,MacLeod2010, Zu2013}. The DRW model yields a variability amplitude, characterised by the standard deviation $\sigma$ of the light curves and a damping timescale $\tau_d$ which represent the thermal timescale of the accretion disk fluctuations. Our calculated value of $\sigma$  is higher than that typically observed in classical Type 1 AGNs. The correlation plot of variability amplitude and the percentage AGN fraction (i.e.,  100 $\times {Flux_{\textsc{agn}}}/{Flux_{\textsc{agn}+\textsc{host}}}$) is shown in the bottom panel of Figure \ref{fig: corr_sigma_host_frac}. There is a minimal correlation between $\sigma$ and the AGN fraction. This is further evident from the median value for our sample is $\log(\sigma) \approx -0.68$, compared to $\log(\sigma) \approx -1.0$ reported for typical NLSy1 and BLSy1 populations in the literature \citep{2017ApJ...842...96R, 2021ApJ...907...96S}.
This increase in amplitude indicates that, even after accounting for a substantial host‐galaxy contribution, the intrinsic variability of low‐mass AGNs remains strong enough to dominate the observed signal.

\begin{figure*} 
    \centering
        
        \includegraphics[width=0.4\textwidth]{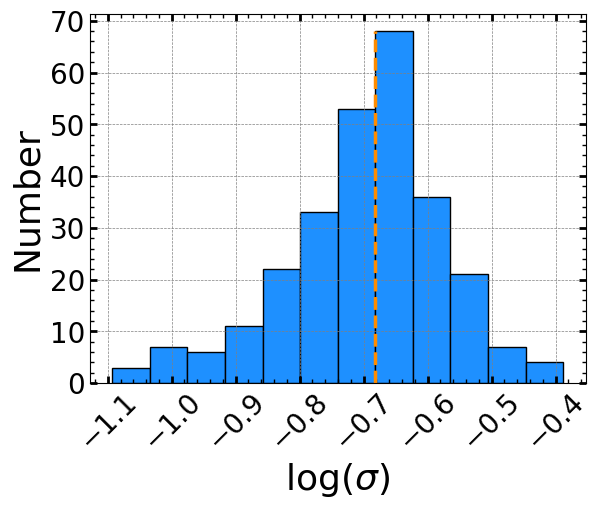}
         \includegraphics[width=0.4\textwidth]{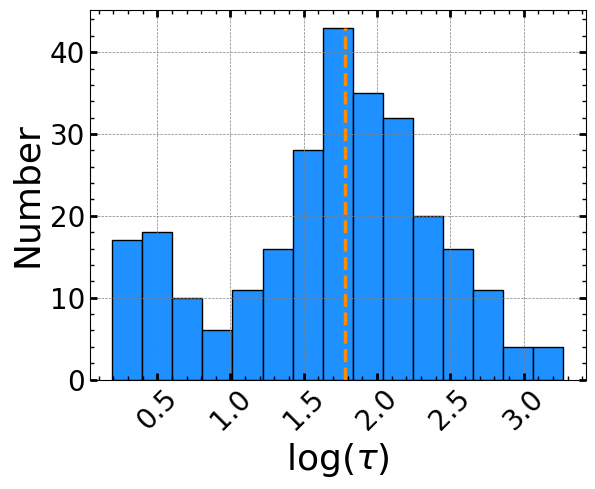}


        \caption{Distributions of the variability parameters obtained from DRW modelling of the $r$-band light curves. The left panel shows the variability amplitude ($\sigma_d$) and the right panel the characteristic timescale ($\tau_d$), with their logarithmic median values of $-0.68$ and $1.78$, respectively.}

        \label{fig: histograms_tau_sigma}
\end{figure*}

The estimated value of damping time scale $\tau_d$ derive from our DRW modeling is shown in right panel of Figure \ref{fig: histograms_tau_sigma} with median value of log ($\tau_d$)  $\approx 1.78$ which corresponds to a characteristic variability timescale of approximately 60 days.
This value is broadly consistent with expectations for low-luminosity AGNs, which are predicted to exhibit shorter thermal or stochastic timescales compared to more massive or higher-luminosity AGNs. 

\section{Discussion } \label{section5}

We studied the physical properties of a sample of 315 low-luminosity AGNs (LLAGNs) hosting low-mass black holes. The study reveals a median \( R_{\text{Fe~{\sc ii}}} \approx 0.61 \), significantly lower than the values reported for NLSy1s, indicating weaker Fe~{\sc ii} emission relative to H\(\beta\). This low \( R_{\text{Fe~{\sc ii}}} \) is consistent with the sub-Eddington accretion rates (median \(\log R_{\text{Edd}} \approx -0.68\)) observed in our sample and suggests that these LLAGNs may operate in a radiatively inefficient accretion regime, such as advection-dominated accretion flows (ADAFs) \citep{1995ApJ...452..710N, yuan_narayan_2014}. The weak Fe~{\sc ii} emission may result from lower ionisation parameters or reduced gas densities in the BLR, conditions less conducive to producing strong Fe~{\sc ii} lines \citep{netzer_2013}. 

The absence of a significant correlation between FWHM H\(\beta\) and \( R_{\text{Fe~{\sc ii}}} \) (\(\rho = 0.07\)) in our sample deviates from the strong anti-correlation observed in luminous quasars, a hallmark of the quasar main sequence (QMS) \citep{borson_greene_1992, shen_ho_2014}. This lack of correlation challenges the universality of the QMS across all AGN luminosities and black hole masses, particularly for LLAGNs with IMBHs. The distinct locus of our sample in the 4DE1 parameter space, characterised by low \( R_{\text{Fe~{\sc ii}}} \) and low \( R_{\text{Edd}} \), suggests that these objects may represent a physically distinct accretion state, potentially driven by different BLR dynamics or accretion disk structures compared to high-luminosity AGNs. These findings align with studies suggesting that low \( R_{\text{Fe~{\sc ii}}} \) is a proxy for lower Eddington ratios, potentially influenced by a weaker soft X-ray excess that limits iron release into the gas phase \citep{2022AN....34310112G}.

From an evolutionary perspective, the BLR in LLAGNs exhibits properties that differ from those in high-luminosity AGNs, reflecting variations in accretion activity and black hole growth. Theoretical models predict that the BLR may disappear in AGNs with luminosities below a critical threshold, estimated at \( L \approx 5 \times 10^{39} (M/10^7 M_{\odot})^{2/3} \) erg s\(^{-1}\), where \( M \) is the black hole mass, due to the transition to radiatively inefficient accretion \citep{2009ApJ...701L..91E,2014MNRAS.438.3340E}. The presence of broad emission lines in our sample and the luminosities indicate that these LLAGNs are above this threshold. However, the weak Fe~{\sc ii} emission suggests that the BLR in these systems may have lower gas densities or ionisation parameters, leading to altered emission-line profiles. Reverberation mapping studies indicate that Fe~{\sc ii} emission originates from the outer BLR, approximately twice the radius of H\(\beta\), consistent with narrower Fe~{\sc ii} line widths \citep{panda_2018,panda_2019a,panda_2019b,2022AN....34310112G}. In LLAGNs, the BLR may be smaller or less extended, reflecting the lower accretion rates and potentially influencing the observed low \( R_{\text{Fe\textsc{ii}}} \).


The IMBHs in our sample, with median masses of \(\log (M_{BH} / M_{\odot}) \approx 5.67\), provide a unique opportunity to explore the evolutionary stages of black hole growth.

While the black hole masses ($\mathrm{M}_{\mathrm{BH}}$) presented in this study are derived from the single-epoch (SE) virial method using the $\mathrm{H}\beta$ broad line and the monochromatic luminosity $\lambda L_{\lambda}(5100\,\text{\AA})$, it is essential to consider their alignment with the fundamental $\mathrm{M}_{\mathrm{BH}}\text{--}\mathrm{M}_{\mathrm{bulge}}$ scaling relation, which reflects the long-term co-evolutionary history of the system. Recent literature focusing on low-mass AGNs and Seyferts  suggests that the slope of this relation may be significantly flatter in the $\mathrm{M}_{\mathrm{bulge}} \lesssim 10^{10}\, \mathrm{M}_{\odot}$ regime than the steep relations derived from local, quiescent, massive elliptical galaxies  \citep[e.g. see][]{2024ApJ...971..173S}. The implication is critical: if we assume the standard, steeper relation holds, our low-$\mathrm{M}_{\mathrm{bulge}}$ systems, possessing $\mathrm{M}_{\mathrm{BH}}$ values tied to the established SE calibration, would appear underestimated relative to that steep benchmark. Conversely, if the flatter slope is correct, the offset represents a genuine physical divergence in the black hole-host growth path, possibly indicative of delayed black hole seeding or varied feedback efficiencies in spiral/disc-dominated hosts.

The low \( R_{\text{Fe~{\sc ii}}} \) and sub-Eddington accretion rates support the hypothesis that these systems are in an early or transitional phase of black hole evolution, potentially bridging the gap between stellar mass and supermassive black holes \citep{volonteri_2010}.

The significant host galaxy contribution, accounting for up to 90\% of the optical flux in the faintest sources, poses a challenge to interpreting LLAGN properties and hence requires careful spectral decomposition.

However, since we are able to model host galaxy contribution in the optical spectra, it has little bearing on the properties studied here, which are derived from the single epoch spectra. Despite this, we now measure a median variability amplitude of $\log\sigma\simeq -0.68$, comparable to or even exceeding typical Type 1 AGNs. This result demonstrates that low‐mass AGNs retain strong intrinsic variability.

In conclusion, our analysis of LLAGNs with IMBHs reveals a distinct population characterised by low \( R_{\text{Fe~{\sc ii}}} \), sub-Eddington accretion, and a deviation from the QMS. These properties suggest a radiatively inefficient accretion regime, likely dominated by ADAFs, and highlight the unique physical conditions within their BLRs. The evolutionary perspective positions these objects as potential seeds or faded remnants of supermassive black holes, offering insights into the early stages of black hole growth. Future multi-wavelength observations, particularly in X-ray and radio bands, are essential to explain further the accretion physics and evolutionary status of these systems.

\vspace{0.5cm}

\section{ Conclusions}
\label{section6}

In this work, we studied a sample of 315 low-luminosity active galactic nuclei (LLAGNs) with low-mass black holes selected from the Sloan Digital Sky Survey (SDSS) Data Release 16, with redshifts $z < 0.35$. By measuring key spectral parameters such as H$\beta$ full-width at half-maximum (FWHM), the Fe~{\sc ii} to H$\beta$ flux ratio (R$_{\text{Fe~{\sc ii}}}$), black hole mass (M$_{\text{BH}}$) and Eddington ratio (R$_{\text{Edd}}$), we have explored the position of these sub-Eddington systems within the quasar main sequence (QMS) framework. Our analysis was also complemented by Zwicky Transient Facility (ZTF) light curves, which address host galaxy contamination and variability. The following key findings summarise our results:

\begin{enumerate}
    \item The median spectral properties of our LLAGN sample, with $H\beta$ FWHM $\approx 992~\text{km}~\text{s}^{-1}$, R$_{\text{Fe~{\sc ii}}} \approx 0.61$, $\log(M_{\text{BH}}/M_{\odot}) \approx 5.67$, and $\log R_{\text{Edd}} \approx -0.68$, indicate sub-Eddington accretion, distinguishing these objects from typical high-luminosity quasars.
    \item We find a strong anticorrelation between R$_{\text{Edd}}$ and H$\beta$ FWHM ($\rho = -0.57$, $p < 10^{-5}$), and no significant correlation between H$\beta$ FWHM and R$_{\text{Fe~{\sc ii}}}$ ($\rho = 0.07$, $p = 0.21$), suggesting diverse accretion dynamics in LLAGNs.
    \item The LLAGN population deviates from the quasar main sequence defined by high-luminosity AGN, occupying a distinct region in the R$_{\text{Fe~{\sc ii}}}$ versus H$\beta$ FWHM plane, likely due to lower accretion rates and significant host galaxy contributions (up to 90\% of optical flux).
    \item Variability analysis using ZTF light curves reveals variability amplitudes of median \(\log\sigma \approx -0.68\), comparable to those of typical Type 1 AGNs, indicating that low-mass AGNs retain substantial intrinsic variability despite significant host-galaxy flux.

  
\end{enumerate}

To summarise, these finding suggest that  LLAGNs with IMBHs reveals a distinct population characterised by low $R_{\text{Fe~{\sc ii}}}$ (median $R_{\text{Fe~{\sc ii}}} \approx 0.61$), sub-Eddington accretion (median $\log R_{\text{Edd}} \approx -0.68$), and a deviation from the QMS. These properties suggest a radiatively inefficient accretion regime, likely dominated by ADAFs, and highlight the unique physical conditions within their BLRs. The evolutionary perspective positions these objects as potential seeds or faded remnants of supermassive black holes, offering insights into the early stages of black hole growth.

These findings highlight the unique spectral and accretion properties of LLAGNs with IMBHs, suggesting that they represent a distinct population within the broader AGN framework. Although limited in number in the SDSS catalogue, the number of such sources is expected to increase significantly in the current and upcoming surveys as the DESI-DR1 survey \citep{2025arXiv250314745D}. Thus, future spectroscopic and variability studies, particularly with high-resolution instruments, will further elucidate their role in galaxy evolution and black hole growth.

\section*{Acknowledgements}

We acknowledge SDSS for making optical spectra publicly available. Funding for the Sloan Digital Sky Survey V has been provided by the Alfred P. Sloan Foundation, the Heising-Simons Foundation, the National Science Foundation, and the Participating Institutions. SDSS acknowledges support and resources from the Center for High-Performance Computing at the University of Utah. SDSS telescopes are located at Apache Point Observatory, funded by the Astrophysical Research Consortium and operated by New Mexico State University, and at Las Campanas Observatory, operated by the Carnegie Institution for Science. The SDSS web site is \url{www.sdss.org}. SDSS is managed by the Astrophysical Research Consortium for the Participating Institutions of the SDSS Collaboration, including Caltech, The Carnegie Institution for Science, Chilean National Time Allocation Committee (CNTAC) ratified researchers, The Flatiron Institute, the Gotham Participation Group, Harvard University, Heidelberg University, The Johns Hopkins University, L'Ecole polytechnique f\'{e}d\'{e}rale de Lausanne (EPFL), Leibniz-Institut f\"{u}r Astrophysik Potsdam (AIP), Max-Planck-Institut f\"{u}r Astronomie (MPIA Heidelberg), Max-Planck-Institut f\"{u}r Extraterrestrische Physik (MPE), Nanjing University, National Astronomical Observatories of China (NAOC), New Mexico State University, The Ohio State University, Pennsylvania State University, Smithsonian Astrophysical Observatory, Space Telescope Science Institute (STScI), the Stellar Astrophysics Participation Group, Universidad Nacional Aut\'{o}noma de M\'{e}xico, University of Arizona, University of Colorado Boulder, University of Illinois at Urbana-Champaign, University of Toronto, University of Utah, University of Virginia, Yale University, and Yunnan University.
 
Based on observations obtained with the Samuel Oschin Telescope 48-inch and the 60-inch Telescope at the Palomar Observatory as part of the Zwicky Transient Facility project. ZTF is supported by the National Science Foundation under Grants No. AST-1440341 and AST-2034437 and a collaboration including current partners Caltech, IPAC, the Oskar Klein Center at Stockholm University, the University of Maryland, University of California, Berkeley, the University of Wisconsin at Milwaukee, University of Warwick, Ruhr University, Cornell University, Northwestern University and Drexel University. Operations are conducted by COO, IPAC, and UW. 

HS and HC express their gratitude to the Inter-University Centre for Astronomy and Astrophysics (IUCAA) for their hospitality and the provision of High-Performance Computing (HPC) facilities under the IUCAA Associate Programme. SP is supported by the international Gemini Observatory, a program of NSF NOIRLab, which is managed by the Association of Universities for Research in Astronomy (AURA) under a cooperative agreement with the U.S. National Science Foundation, on behalf of the Gemini partnership of Argentina, Brazil, Canada, Chile, the Republic of Korea, and the United States of America.

\section*{Data Availability}

The data used in this study are publicly available from the Sloan Digital Sky Survey Data Release 16 (https://www.sdss.org/dr16/) and the Zwicky Transient Facility (https://www.ztf.caltech.edu/). Derived parameters and spectral fits are available upon request.


\bibliographystyle{mnras}
\bibliography{main} 

\clearpage
\appendix

\onecolumn
\section{Example of poorly fitted sources.} \label{appendixA}
In this section, we illustrate a handful of spectra where the broad $H\beta$ component is either weak, absent, or the signal-to-noise ratio is too low for obtaining a reliable fit. These sources have been omitted from our sample for further analysis.


\begin{center}
        \includegraphics[width=17cm, height=9cm]{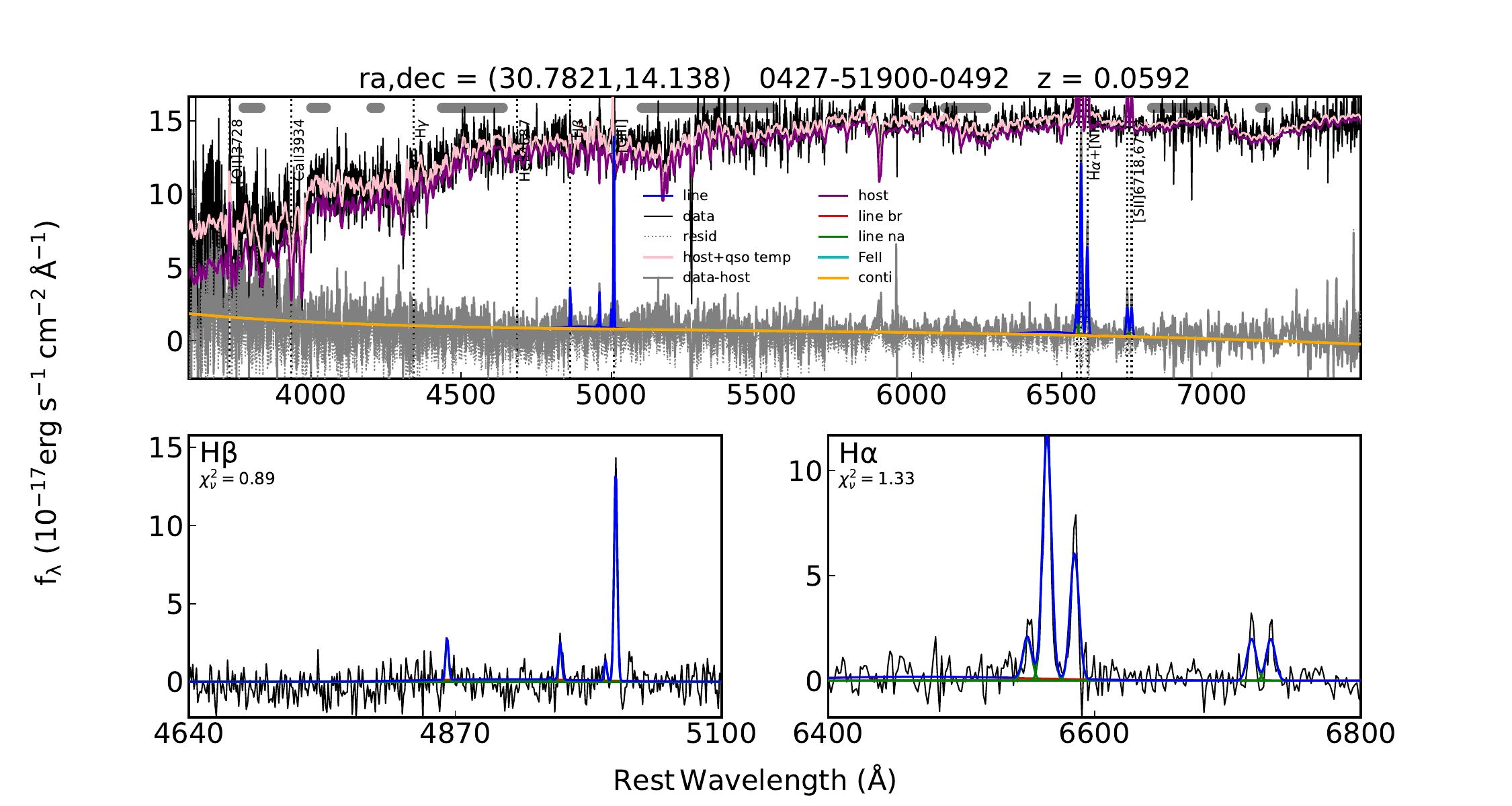} 
          \includegraphics[width=17cm, height=9cm]{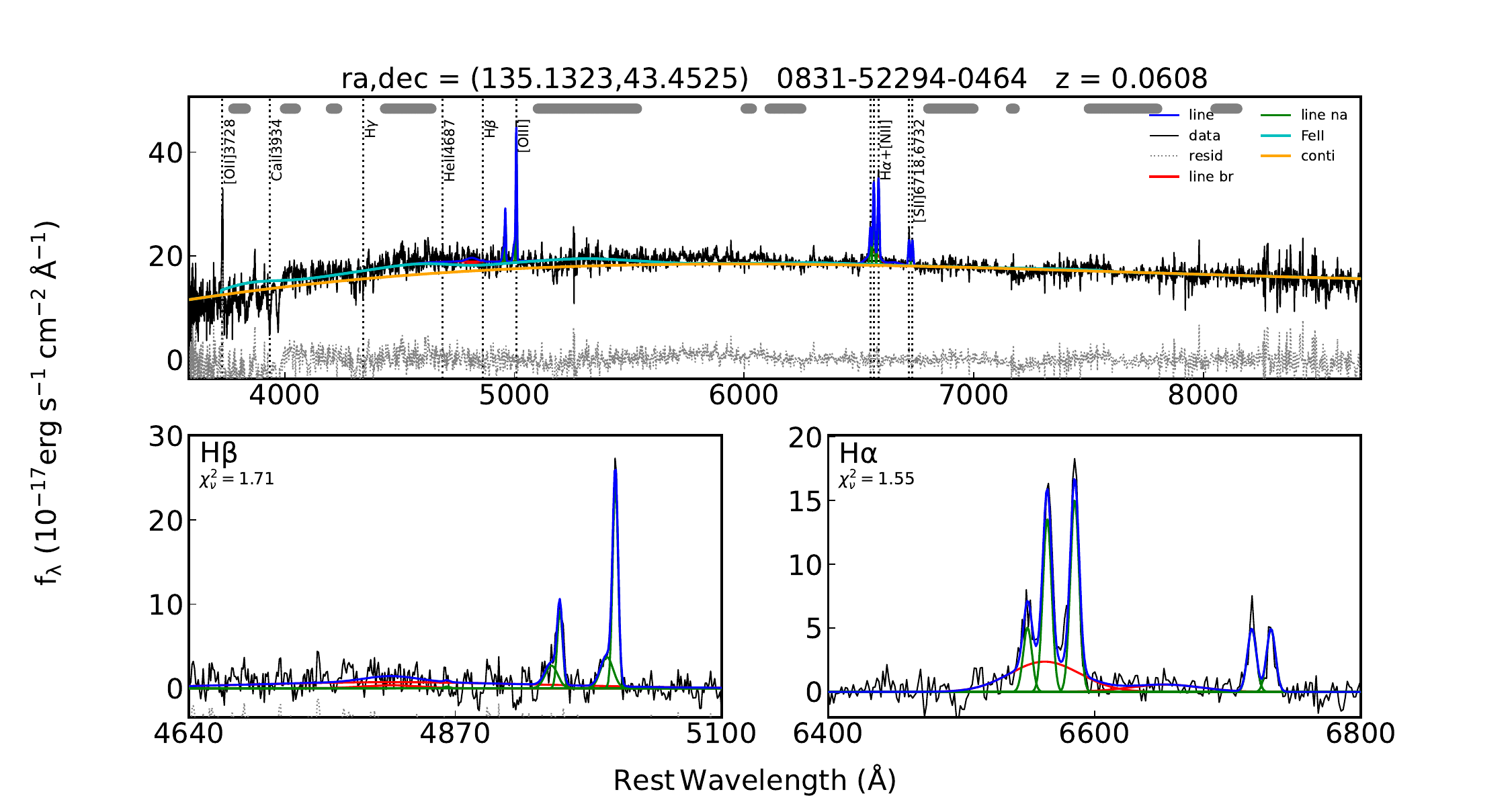}
        \captionof{figure}{The sources in this plot have no detectable $H\beta$ emission line; hence, they could not be analysed properly.}
        \label{fig:a1}
\end{center}  

\clearpage
\onecolumn
\begin{center}
\ 
      
        \includegraphics[width=17cm, height=9cm]{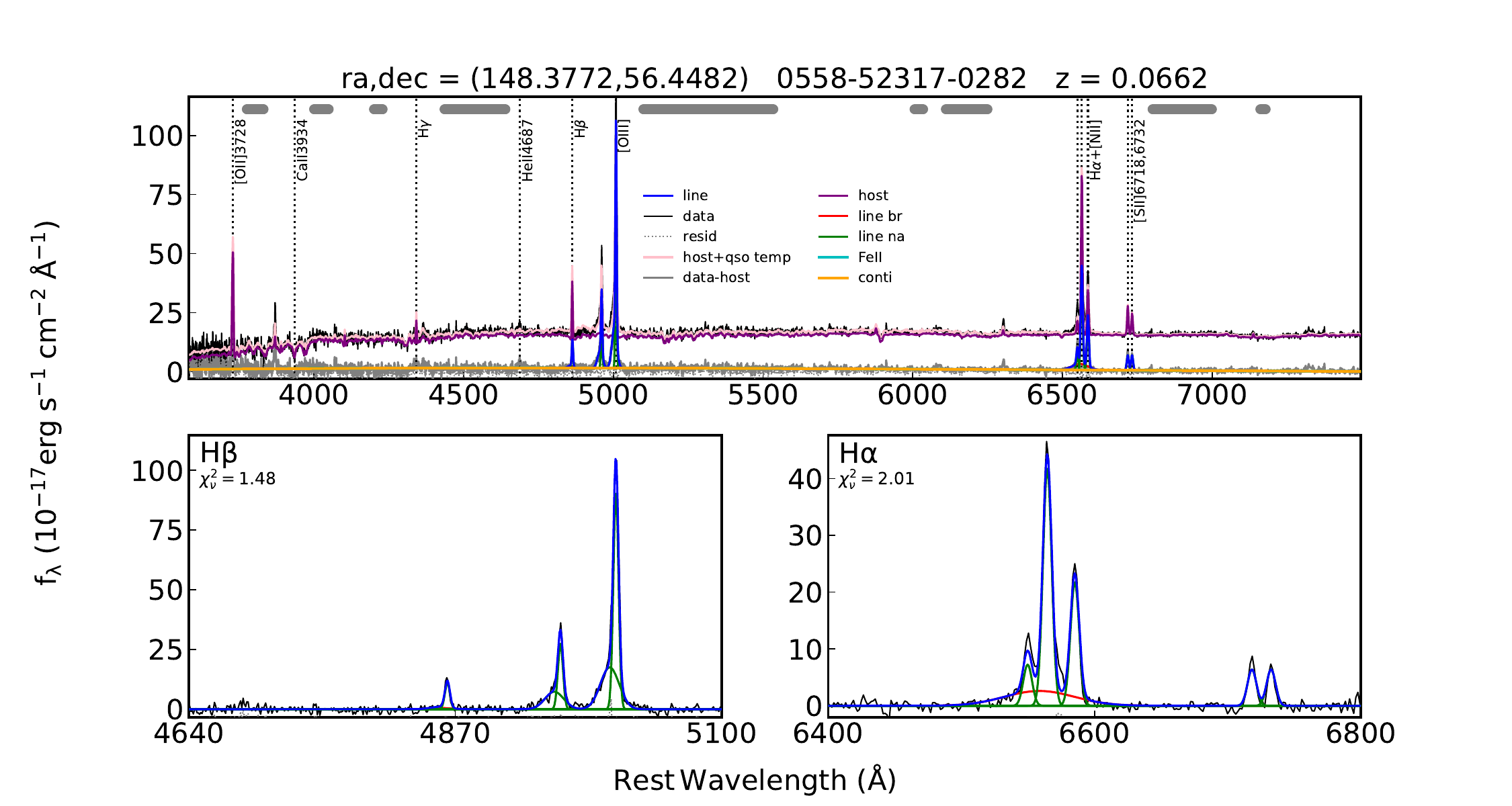}
        \includegraphics[width=17cm, height=9cm]{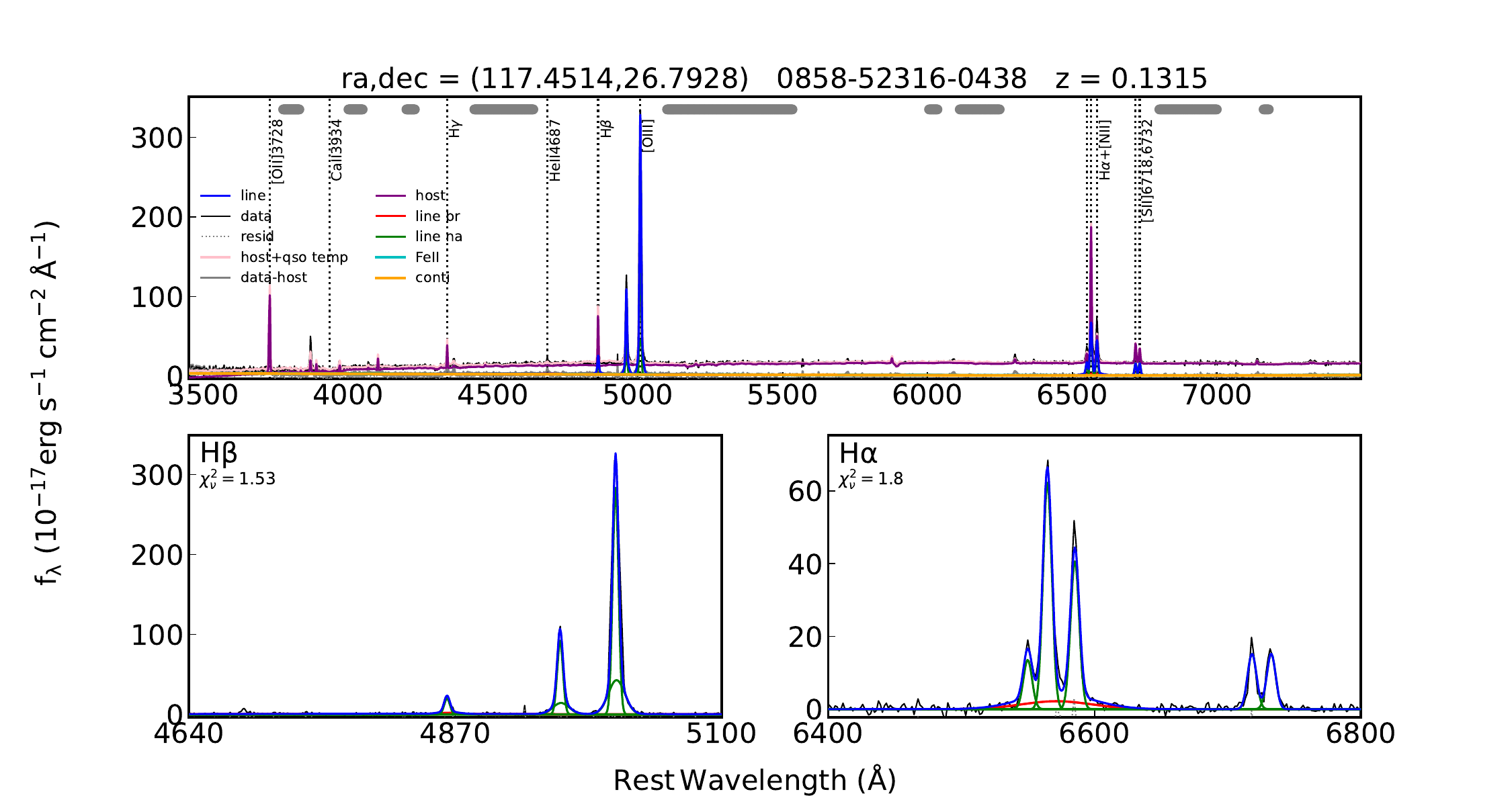}
        \captionof{figure}{The sources in this plot, even though they have a feeble $H\beta$ emission visible, the broad component could not be calculated, and they could not be analysed properly.}
        \label{fig:a2}
\end{center}


\bsp	
\label{lastpage}
\end{document}